# Nonlinear Mid-infrared Metasurface based on a Phase-Change Material

Fuyong Yue[1, †], Riccardo Piccoli[1, †], Mikhail Y. Shalaginov[2, †], Tian Gu[2, 3, *], Kathleen Richardson[4], Roberto Morandotti[1, 5], Juejun Hu[2], and Luca Razzari[1, *]

[1]INRS-EMT, 1650 Boulevard Lionel-Boulet, Varennes, Québec J3X 1S2, Canada

[2]Department of Materials Science & Engineering, Massachusetts Institute of Technology, Cambridge, MA, USA

[3]Materials Research Laboratory, Massachusetts Institute of Technology, Cambridge, MA, USA

[4]The College of Optics & Photonics, Department of Materials Science and Engineering, University of Central Florida, Orlando, FL, USA

[5]Institute of Fundamental and Frontier Sciences, University of Electronic Science and Technology of China, Chengdu 610054, Sichuan, China

*Email: gutian@mit.edu
*Email: razzari@emt.inrs.ca
† These authors contributed equally to this work.

**Abstract**

**The mid-wave infrared (MWIR) spectral region (3-5 μm) is important to a vast variety of applications in imaging, sensing, spectroscopy, surgery, and optical communications. Efficient third-harmonic generation (THG), converting light from the MWIR range into the near-infrared, a region with mature optical detection and manipulation technologies, offers the opportunity to mitigate a commonly recognized limitation of current MWIR systems. In this work, we present the possibility of boosting THG in the MWIR through a metasurface design. Specifically, we demonstrate a 30-fold enhancement in a highly nonlinear phase change material $Ge_2Sb_2Se_4Te_1$ (GSST), by patterning arrays of subwavelength cylinders supporting a magnetic dipolar resonance. The unprecedented broadband transparency, large refractive index, and remarkably high nonlinear response, together with unique phase-change properties, make GSST-based metasurfaces**

an appealing solution for reconfigurable and ultra-compact nonlinear devices operating in the MWIR.

**Introduction**

The mid-wave infrared (MWIR), ranging from 3 to 5 μm wavelengths[1], coincides with the primary vibrational absorption bands of many molecules (e.g. carbon monoxide (CO) and methane ($CH_4$)), as well as the atmospheric transmission window. It is therefore of crucial importance for a vast variety of applications, including optical imaging[2,3], sensing[4], spectroscopy[5], surgery[6], and free-space optical communications[7]. However, the development of MWIR systems has considerably lagged behind advances realized by visible or near-infrared (NIR) counterparts, due to the complexity and cost of optical and optoelectronic components in this wavelength range.

Optical metasurfaces composed of artificial subwavelength building blocks, i.e. "meta-atoms", offer a potential solution to this challenge by enabling unprecedented control over the electromagnetic response of materials, thereby breaking the bounds traditionally imposed by their chemical composition[8-10]. A number of metasurface-based MWIR elements operating in the linear regime, such as flat lenses and polarization controllers, have been fabricated out of metals[11] and high-index semiconductors[12]. Metasurfaces also provide a powerful means to engineer the nonlinear response of materials, as has been largely shown thus far at shorter wavelengths[13-26]. Indeed, the effective nonlinear polarization of metasurfaces can be tailored and boosted by resonance-induced electromagnetic field enhancement[13-17]. Plasmonic metasurfaces made of noble metals such as gold and silver have been initially employed to enhance nonlinear processes such as second-harmonic generation[18], THG[19], and four-wave mixing[20]. However, the overall conversion efficiency of plasmonic devices has been limited by high absorption losses and relatively low damage thresholds[13]. To overcome these limitations, high-index low-loss dielectric metasurfaces have emerged as a promising platform. Semiconductor materials, such as Ge and Si, have been successfully employed to realize dielectric resonators for enhancing THG in the visible/NIR[21] and boosting high-harmonic generation in the ultraviolet[22]. In addition, semiconducting metasurfaces featuring second-order nonlinearity have also been investigated for second-harmonic generation[23,24]. A significant variety of resonance-based approaches[15,21,25,26] have been extensively exploited to enhance nonlinear wave mixing processes in such systems.

Frequency conversion from the MWIR to the NIR (1-1.7 μm) circumvents the challenge associated with MWIR light manipulation and detection. Related technologies have already seen applications in nonlinear holography and imaging[27], mid-infrared all-optical signal processing[28], and in particular frequency upconversion imaging (FUI)[3,29]. For instance, FUI technology enables mid-infrared imaging using standard visible and NIR cameras, leveraging THG or sum frequency generation as the upconversion mechanism. In this scenario, strongly-nonlinear optical materials with broadband infrared transparency are highly demanded. For this purpose, chalcogenide alloys[30] represent a promising class of materials since they feature a broad transparency window, large Kerr nonlinearity, and unique phase change characteristics. In particular, phase transition allows the additional possibility to dynamically control the nonlinearity, thus introducing an important degree of freedom.

The archetypal phase change alloy $Ge_2Sb_2Te_5$ (GST) has been widely used in active linear metasurfaces due to its dramatic optical property contrast between the amorphous and crystalline phases[31-35]. More recently, a new phase change material $Ge_2Sb_2Se_4Te_1$ (GSST), the Se-substituted counterpart of GST, has been identified with a similarly large index contrast, while claiming unique bi-state transparency over an extremely broad range of frequencies[36-39]. This unique property poises GSST as a superior optical phase change material for implementing nonvolatile active optical components. Recent works have demonstrated several linear optical devices based on GSST, including an integrated photonic switch[37], a spatial light modulator[37], a tunable metasurface filter[39], and a reconfigurable varifocal metalens[36]. In nonlinear optical applications, one can envision that the nonlinear response of this phase change material can be equivalently manipulated via external (e.g., electrical, optical, or thermal) stimuli.

In this paper, we have explored the nonlinear optical performance of the phase change material GSST, targeting THG with the fundamental frequency in the MWIR range. The third-order nonlinear susceptibilities $\chi^{(3)}$ of amorphous and crystalline GSST films have been evaluated from the THG conversion efficiency. Subsequently, we have demonstrated an all-dielectric nonlinear metasurface for the MWIR, by patterning a properly-designed array of crystalline GSST meta-atoms and thus achieving a significant enhancement of the THG efficiency. The critical advantages of this device include the strong nonlinear light-matter interaction deriving from both the material properties and the device architecture, broadband transparency, and the possibility of modulating the nonlinearity via material phase transition. For these reasons, GSST-based metasurfaces represent a solid step toward the next generation of nonlinear metadevices with reconfigurable properties.

# Results

**Characterization of the nonlinear optical properties of crystalline and amorphous GSST films.** One-micron-thick amorphous and crystalline GSST films were deposited on CaF$_2$ substrates for optical characterization. The details of sample fabrication can be found in **Methods**. To investigate the nonlinear performance of the phase change material, we exposed the amorphous and crystalline GSST films to a high intensity MWIR laser beam and characterized the THG response. Figure 1 (a) shows the schematic of the experimental setup, consisting of a tunable MWIR femtosecond laser, an objective lens (NA = 0.85, M-60X - *Newport*) to collect the THG signal, a monochromator, and a camera. To quantify the THG conversion efficiency, instead of a camera, we used a Ge-based high-gain detector coupled to a lock-in amplifier. More details about the experimental setup can be found in **Methods**. The THG spectra produced by the crystalline GSST film at different pump wavelengths are shown in Fig. 1 (b). The observed bandwidth ($\Delta\lambda_{THG} \approx 40$ nm) is consistent with THG pulses generated by our MWIR femtosecond pump laser. Figure 1(c) shows the measured external power-to-power conversion efficiency $\eta_P^{ext}$ (i.e., the ratio of the THG power exiting the sample to the incident pump power) as a function of the pump intensity, highlighting the expected quadratic dependence. The inset of the figure reports the observed beam profile of the generated THG under illumination of the pump laser at a wavelength of 4.36 μm. Importantly, the crystalline GSST film was robust to high-intensity MWIR femtosecond pulses, as confirmed by reproducible and reversible THG spectra and power trends (see Section 1 of Supporting Information). Overall, the nonlinear response of both unpatterned films, and the metasurface investigated below, was found to be stable under all the reported experimental conditions. No sign of visual degradation was observed for all the samples.

One of the key appealing features of the proposed GSST platform for nonlinear applications lies in the possibility of actively switching the system nonlinearity via phase transition, allowing the opportunity to dynamically modulate the conversion efficiency. To retrieve the third-order susceptibilities of GSST films and investigate such nonlinear optical contrast, the THG efficiency was quantified for the two structural states. The THG power is expected to increase along the propagation direction of the fundamental beam inside the nonlinear medium, until a maximum is achieved at the coherence length $L_{coh} = \frac{\lambda_p}{|6n_\omega - 6n_{3\omega}|}$, where $\lambda_p$ is the vacuum wavelength of the pump, while $n_\omega$ and $n_{3\omega}$ are the refractive indices of the probed material at the fundamental and THG wavelengths, respectively[40]. In our case, the thickness of the nonlinear medium is $L \approx 1$ µm, while $L_{coh} \geq 1.7$ µm. The effective value of the third-order susceptibility $\chi^{(3)}$ can be retrieved from the following formula:

$$\chi^{(3)} = \frac{2\epsilon_0 c \lambda_p \sqrt{n_\omega^3 n_{3\omega}}}{\sqrt[4]{3}\pi L I_P^{int} sinc\left(\frac{\Delta k L}{2}\right)} \sqrt{\eta_P^{int}}, \quad (1)$$

where $I_P^{int}$ is the internal pump peak intensity (evaluated in the centre of the beam at its waist), $\epsilon_0$ is the vacuum permittivity, $\Delta k$ is the wavevector mismatch, $c$ is the speed of light, and $\eta_P^{int} = \frac{P_{3\omega}}{P_\omega}$ is the THG internal power-to-power conversion efficiency (see Section 2 of the Supporting Information for definitions and detailed derivations).

The measured $\eta_P^{ext}$ of the crystalline and amorphous GSST films at the fundamental wavelength of 4.5 µm (average pump power of 10.7 mW, corresponding to an intensity of 0.34 GW/cm$^2$) are $(2.41 \pm 0.32) \times 10^{-9}$ and $(2.71 \pm 0.74) \times 10^{-10}$, respectively. Consequently, using Eq. 1, we estimated the corresponding values of the third-order susceptibilities of the two GSST films to be $\chi_c^{(3)} = (3.36 \pm 0.41) \times 10^{-18}$ $(m^2/V^2)$ and $\chi_a^{(3)} = (4.58 \pm 0.59) \times 10^{-19}$ $(m^2/V^2)$. The noticeable difference between the $\chi^{(3)}$ values in the two phases (a contrast ratio of ~7.4) makes GSST an appealing material for developing reconfigurable nonlinear optical devices.

**Table I**: Values of the refractive indices, conversion efficiencies, experimental and Miller's estimated $\chi^{(3)}$ for the crystalline (C-GSST) and amorphous (A-GSST) films.

|  | $n_\omega$ | $n_{3\omega}$ | $\eta_P^{ext} \times 10^{-9}$ | Experimental $\chi^{(3)} \times 10^{-18}$ $(m^2/V^2)$ | Miller's estimation $\chi^{(3)} \times 10^{-18}$ $(m^2/V^2)$ |
|---|---|---|---|---|---|
| C-GSST | 4.08 ± 0.08 | 4.51 ± 0.09 | 2.41 ± 0.32 | 3.36 ± 0.41 | 3.50 |
| A-GSST | 3.25 ± 0.06 | 3.38 ± 0.06 | 0.271 ± 0.074 | 0.458 ± 0.059 | 0.468 |

To further corroborate our measured data, we utilized the so-called Miller's rule, which allows evaluating $\chi^{(3)}$ from the linear refractive index $n$[41]. As summarized in Table I, the estimates obtained following this procedure are in very good agreement with our experimental results. To provide additional insight into the nonlinear optical performance of GSST, we compared its third-order susceptibilities with those of other representative nonlinear optical materials in the same spectral region (Fig. 1 (d)). To this end, we employed the generalized Miller's formula[42] to project the $\chi^{(3)}$ values reported in literature at the same set of wavelengths of our experiment ($\lambda_p$ = 4.5 µm, $\lambda_{THG}$ = 1.5 µm). The details about the derivation can be found in Section 3 of the Supporting Information. Figure 1 (d) shows the relation between $\chi^{(3)}$ and $n$ (at 4.5 μm) according to the Miller's rule (dashed line), as well as the $\chi^{(3)}$ values of the GSST films and of the other nonlinear media. The plot clearly indicates that GSST emerges as a high refractive index material with strong phase-dependent nonlinearity.

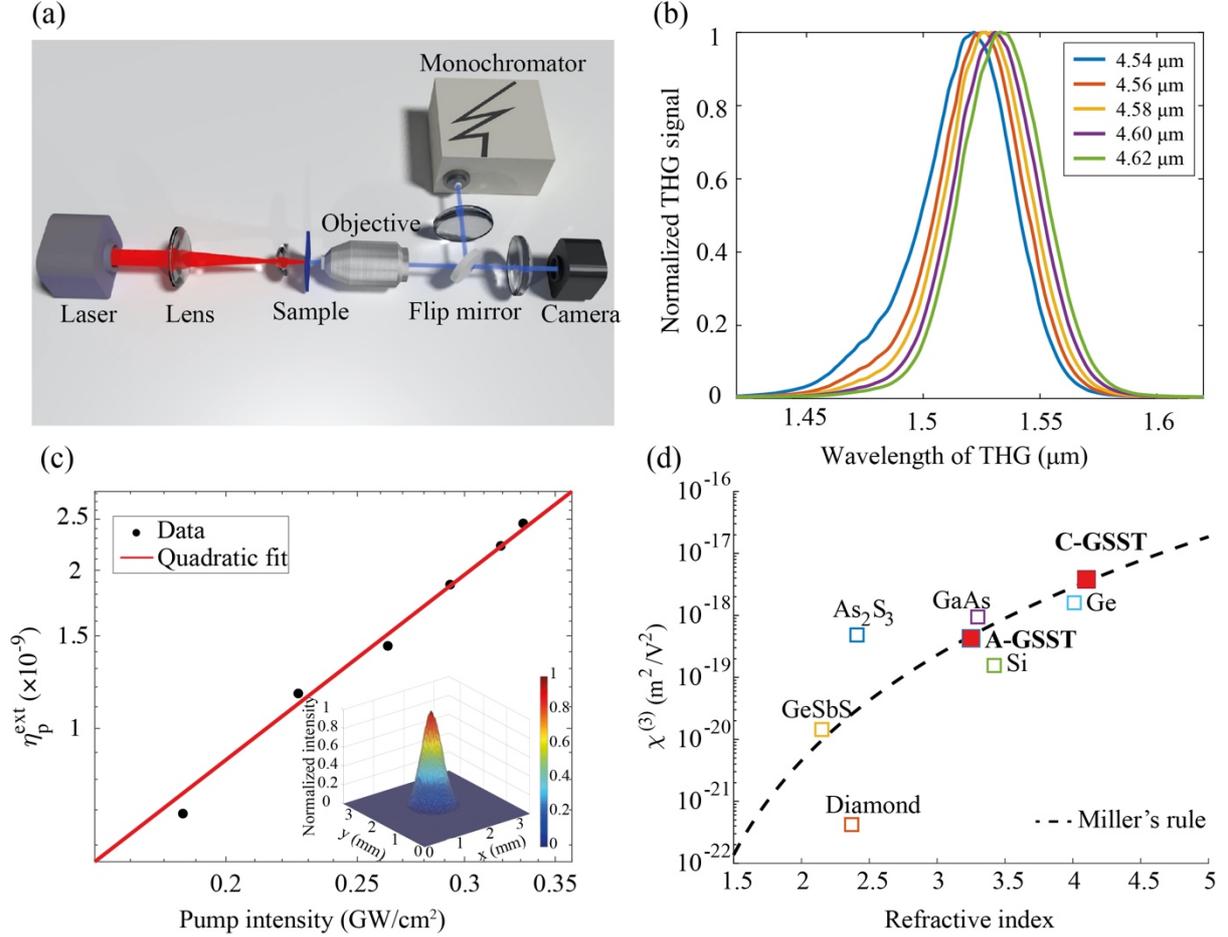

**Figure 1. Experimental characterization of THG in GSST films.** (a) Schematic of the experimental setup. (b) Normalized THG spectra from the crystalline GSST film at different pump wavelengths. (c) Measured conversion efficiency (in log scale) for the crystalline GSST film at a wavelength of 4.36 μm. Red line: quadratic fit. Black dots: measured data. Inset: intensity profile of the THG light captured by the InGaAs camera. (d) $\chi^{(3)}$ as a function of the linear refractive index at the fundamental wavelength of 4.5 μm. The dashed line indicates the Miller's rule prediction. The squares denote the $\chi^{(3)}$ values of various nonlinear materials relevant for the mid-infrared range (nanocrystalline diamond, $\chi^{(3)} = 4.2 \times 10^{-22}$ $(m^2/V^2)$; Si, $\chi^{(3)} = 1.5 \times 10^{-19}(m^2/V^2)$; Ge, $\chi^{(3)} = 1.6 \times 10^{-18}(m^2/V^2)$; As$_2$S$_3$, $\chi^{(3)} = 4.8 \times 10^{-19}(m^2/V^2)$, GaAs, $\chi^{(3)} = 9.5 \times 10^{-19}(m^2/V^2)$, Ge$_{23}$Sb$_7$S$_{70}$ (GeSbS), $\chi^{(3)} = 1.4 \times 10^{-20}(m^2/V^2)$),[40, 43-46], as well as the values for the crystalline (C-GSST) and amorphous (A-GSST) films experimentally evaluated in this work.

**THG enhancement in a GSST metasurface.** Patterning high-index dielectric films with sub-wavelength structures can significantly boost the nonlinear frequency conversion efficiency[22,23,25,47]. This enhancement effect is associated with the tight field confinement of the pump inside the meta-atoms supporting electric/magnetic dipolar resonant modes. The implemented metasurface consisted of crystalline-GSST cylinder meta-atoms arranged in a square lattice. We targeted the exploitation of the magnetic dipolar mode that can be excited inside the meta-atoms due to the circular displacement of a "virtual current" induced by the electric field propagating through the volume[48]. The resonant wavelength of the magnetic mode can be tuned by adjusting the geometry of the cylinders. The dry etching process used in metasurface fabrication causes a slightly non-vertical sidewall profile with a sidewall angle of approximately 85°, which was considered in the numerical simulations. The geometric parameters of the designed metasurface are shown in Fig. 2 (a). The simulated reflection and transmission of the GSST cylinder array situated on a $CaF_2$ substrate depict two pronounced resonances at 3.6 μm and 4.35 μm (Fig. 2 (b)). More details about the simulations can be found in **Methods**. The simulated Poynting vector distributions shown in Fig. 2 (c) and (e) confirm that the two transmission dips correspond to the electric ($\lambda$ = 3.6 μm) and magnetic ($\lambda$ = 4.35 μm) dipole modes, respectively. The simulated electric and magnetic field maps at 4.35 μm (Fig. 2 (d) and (f)) show clear field enhancement within the cylinder volume, as a direct consequence of the magnetic dipole resonance.

The metasurface was fabricated by depositing a one-micron-thick crystalline GSST film on a $CaF_2$ substrate and subsequently patterning it with electron beam lithography followed by reactive ion etching (see **Methods** for further details). Figure 3 (a) shows scanning electron microscopy (SEM) images of the fabricated metasurface and a zoomed-in scan of a single cylinder. First, we characterized the metasurface by measuring its transmittance under the illumination of a nonpolarized broadband light source in the wavelength range of 2-5 μm (VERTEX 70 - *Bruker*). As seen from Fig. 3 (b), the measured transmittance (black dashed curve) confirms the resonance spectral positions predicted by the numerical simulations (red curve).

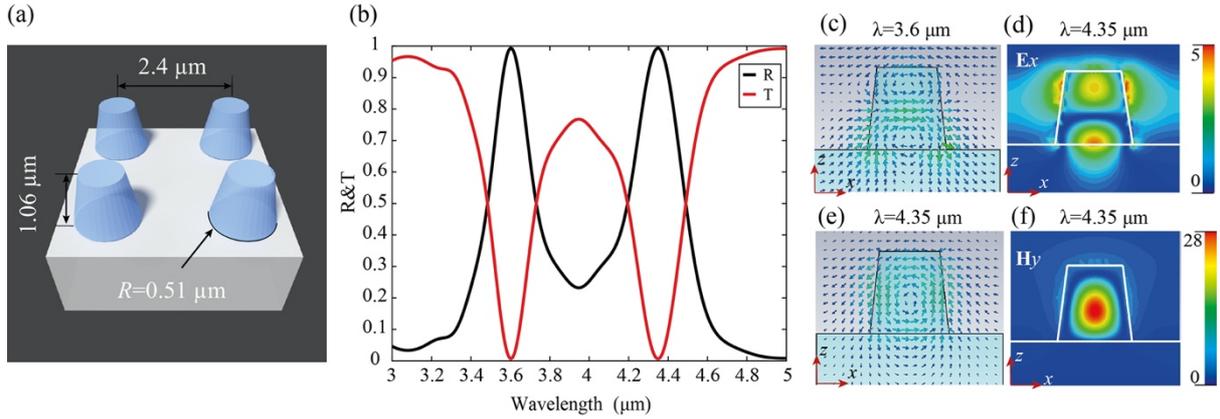

**Figure 2. Design of the GSST metasurface.** (a) Illustration of the metasurface for enhancing THG at the fundamental wavelength of 4.35 μm. Metasurface geometry: lattice constant - 2.4 μm, cylinder height - 1.06 μm, cylinder bottom radius - 0.51 μm, sidewall angle - 85°. (b) Simulated transmittance and reflectance of the GSST metasurface situated on a $CaF_2$ substrate. Poynting vector maps in the *x-z* plane at a wavelength of 3.6 μm (c) and 4.35 μm (e), respectively. Electric (d) and magnetic (f) field components within a GSST cylinder at the wavelength of 4.35 μm. The colour bars indicate the corresponding field enhancement factors normalized to the incident field. All simulations were performed under x-polarized plane wave excitation using CST Microwave Studio®.

During the nonlinear optical characterization, the metasurface was illuminated at normal incidence from the substrate side and the transmitted THG signal was collected with the microscope objective (see the schematic in Fig. 1 (a)). Since the square lattice period of the metasurface (2.4 μm) is larger than the THG wavelength, multiple diffraction orders were observed in the far-field, resulting in 9 THG beams of approximately equal intensity (see sketch in Fig. 3 (c), camera image in Fig. 3 (d), as well as Section 6 of the Supporting Information for a quantitative evaluation). Such structured THG pattern directly shows a straightforward example of the possibilities offered by nonlinear metasurfaces in shaping the emission of the generated light. The metasurface enhancement factor was calculated as a ratio of the overall THG power produced by the metasurface (see Section 6 of the Supporting Information for its estimation procedure) to the THG power from the unpatterned crystalline film of the same thickness, under identical illumination conditions. As can be seen from Fig. 3 (e), the spectral dependence of the THG enhancement arising from the magnetic dipolar resonance perfectly matches the measured far-field transmittance dip. A maximum THG enhancement factor of 28 was estimated at 4.36 μm. It is worth stressing that, in our estimation, only the THG light emitted in the forward direction was considered, while THG is expected to occur in both forward and backward directions[48]. By properly engineering the interaction between the electric and magnetic dipole modes, unidirectional scattering can be realized, which can further increase the overall efficiency of the nonlinear metasurface[49]. Finally, the THG conversion efficiency of the metasurface was characterized at the resonant wavelength of 4.36 μm. As shown in Fig. 3 (f), such efficiency exhibits the expected quadratic dependence with respect to the pump intensity until a deviation from this trend occurs above 0.3 GW/cm$^2$. Such behaviour is commonly observed in nonlinear dielectric metasurfaces[13] and may be related to the appearance of other competing nonlinear processes, such as multi-photon and free-carrier absorption. In our experiments, we observed conversion efficiencies for the GSST metasurface at $\lambda_p$=4.36 μm up to $4.6 \times 10^{-7}$, corresponding to a THG average power of 3.7 nW for an average pump power of 8 mW. In comparison to a previously reported Si-disk-based nonlinear metasurface supporting a magnetic dipolar resonance and operating at shorter wavelengths (conversion efficiency $\eta \approx 10^{-7}$ at a wavelength of 1.24 μm and pump peak intensity of 5.5 GW/cm$^2$)[25], the GSST metasurface achieves an almost 5 times higher conversion efficiency for a pump intensity that is more than 4 times lower (1.23 GW/cm$^2$ in our case).

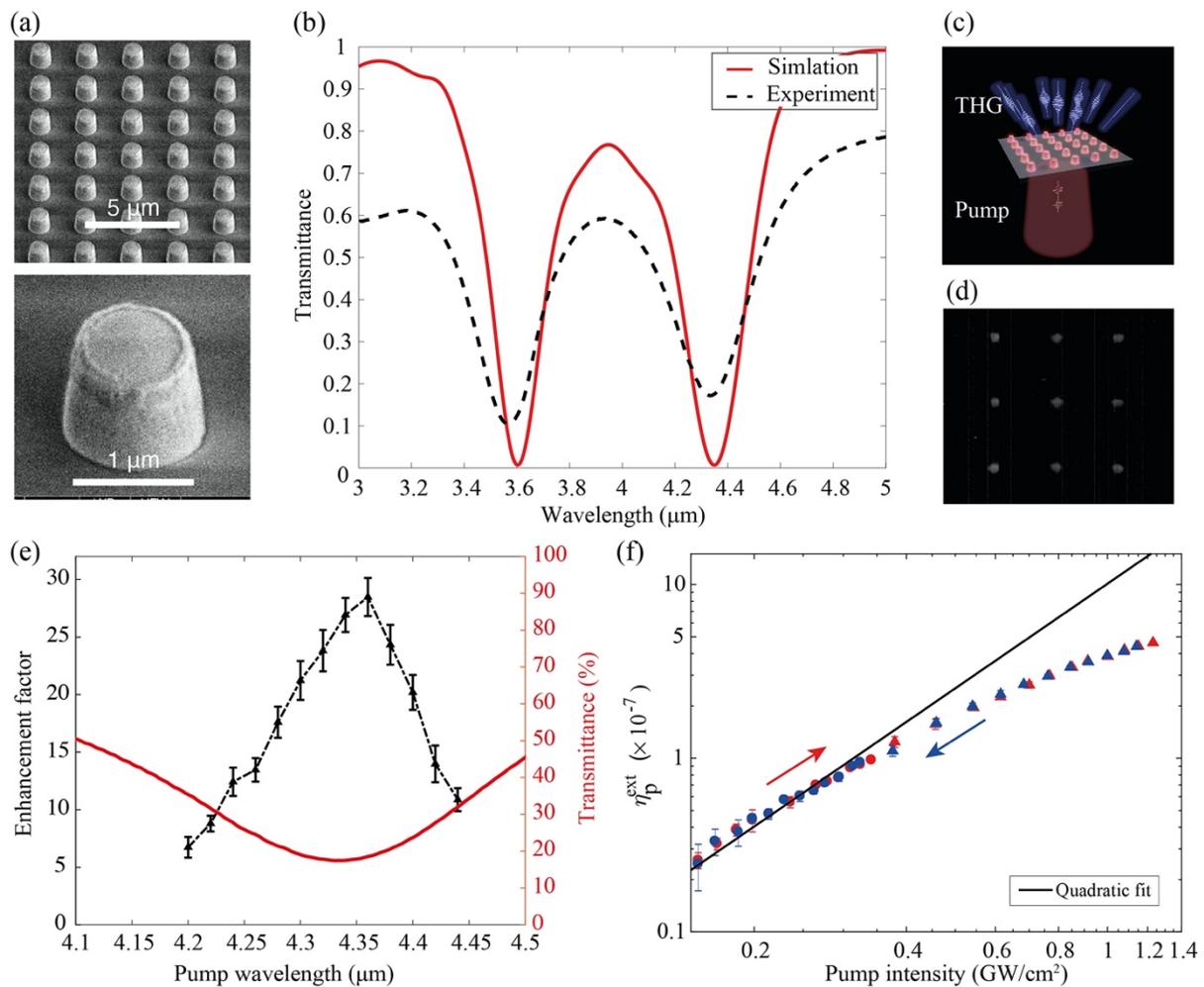

**Figure 3. Linear and nonlinear characterization of the GSST metasurface.** (a) SEM images of the fabricated crystalline-GSST metasurface, with an enlarged image of an individual meta-atom. (b) Simulated and measured (FTIR) transmittance of the GSST metasurface situated on a CaF$_2$ substrate.

(c) Schematic of the THG process in the GSST metasurface. (d) Experimentally recorded image of the diffraction pattern of THG light for a pump wavelength of 4.2 μm. (e) Measured THG enhancement factor (normalized to the THG power from the unpatterned crystalline-GSST film averaged over the considered spectral range) at a pump intensity of 0.34 GW/cm$^2$ (black line), which results to be well-aligned to the measured transmittance dip of the magnetic dipole mode (red line). (f) Measured conversion efficiency $\eta_P^{ext}$ of the metasurface as a function of the pump intensity, at the pump wavelength of 4.36 μm; black line: quadratic fit. Circles and triangles (red for increasing pump intensity and blue for decreasing pump intensity) denote the measured conversion efficiencies for a beam waist of 230 μm and 105 μm, respectively.

## Conclusion

In this paper, we have shown that a significantly enhanced THG in the MWIR can be realized by means of a metasurface design. In particular, we have demonstrated a 30-fold enhancement of the conversion efficiency by patterning a highly nonlinear GSST crystalline film into arrays of meta-atoms supporting a magnetic dipolar resonance. The retrieved effective value of $\chi_c^{(3)}$ for the unpatterned GSST film in the crystalline state at the pump wavelength of 4.5 μm has been found to be $(3.36 \pm 0.41) \times 10^{-18}$ $(m^2/V^2)$, about 4 orders of magnitude larger than that of diamond[44], and ~7 times higher than that of As$_2$S$_3$ chalcogenide glass[45]. For the GSST film in the amorphous state, $\chi_\alpha^{(3)} = (4.58 \pm 0.59) \times 10^{-19}$ $(m^2/V^2)$, revealing a noticeable change in the nonlinear response of the material between the two phases. The marked contrast in the refractive indices and third-order susceptibilities of the two phases makes this material promising for novel, dynamically reconfigurable nonlinear optical architectures. It is important to underline that the phase transition in the GSST meta-atoms not only modifies their effective nonlinear susceptibility, but also switches on/off the microstructure dipolar resonances and the associated field enhancement via the modulation of the linear refractive index. This in turn enables an additional boost in the achievable modulation of the generated THG power via phase transition (more than a two order of magnitude difference in our experimental conditions). Furthermore, metasurfaces based on this new phase change material give access to wavefront manipulation on a subwavelength scale for the nonlinearly-generated light, thus opening new avenues for realizing active mid-infrared nonlinear devices with unprecedented functionalities (encompassing, for example, reconfigurable metalenses for upconversion imaging, nonlinear holography and encryption, just to name a few[50-52]).

## Methods

**Material synthesis and characterization**

The GSST films were deposited onto 1 cm$^2$ double-side polished CaF$_2$ (111) substrates (MTI Corp) by thermal co-evaporation in a custom-made system (PVD Products Inc)[53]. The desired material stoichiometry was achieved by controlling the ratio of evaporation rates of the isolated targets of Ge$_2$Sb$_2$Te$_5$ and Ge$_2$Sb$_2$Se$_5$. The combined deposition rate was kept at approximately 16 Å/s at a base pressure of $3 \times 10^{-6}$ Torr. During the deposition process the sample holder was kept at nearly room temperature and rotated at 6 rpm. The as-deposited GSST film was in its amorphous state. To obtain the crystalline state, the film was annealed on a hot plate at 250°C for 1 hour in argon atmosphere. The film thicknesses were measured to be 1.10 μm and 1.06 μm in amorphous and crystalline states, respectively. The GSST film experienced volumetric contraction of 4% during crystallization. The material stoichiometry was verified by wavelength dispersive spectroscopy (WDS) on a JEOL JXA-8200 SuperProbe Electron Probe Microanalyzer (EPMA). Refractive indices of the amorphous and crystalline GSST films were retrieved using variable angle spectroscopic ellipsometry (J.A.Wollam) combined with the fitting of transmittance/reflectance FTIR data[37].

**Fabrication of the GSST metasurface**

Patterning of the film was performed by electron beam (e-beam) lithography (Elionix ELS F-125) followed by reactive ion etching (Plasmatherm, Shuttlelock System VII SLR-770/734). E-beam writing was done on an 800-nm-thick layer of ZEP520A photoresist, which was spin-coated on top of the GSST film at 2000 rpm for 1 min and afterwards baked at 180 °C for 1 min. Before coating photoresist, the sample surface was treated with standard oxygen plasma cleaning. To suppress charging effects during the e-beam writing process, the photoresist was covered with a water-soluble conductive polymer (Espacer 300Z, Showa Denko America, Inc.). The metasurface pattern was written with the following e-beam settings: voltage of 125 kV, aperture diameter of 120 μm, current of 10 nA, proximity error correction with a base dose time of 0.03 μs/dot (which corresponds to a dosage of 300 μC/cm$^2$). The exposed photoresist was developed by subsequently dipping the sample into water, photoresist developer (ZED-N50), methyl isobutyl ketone (MIBK), and isopropanol alcohol (IPA) for 1 min each. Reactive ion etching was conducted in a gas mixture of CHF$_3$:CF$_4$ (3:1) with respective flow rates of 45 sccm and 15 sccm, pressure of 10 mTorr, and RF power of 200 W. The rate of etching was approximately 80 nm/min. The etching was done in three cycles of 5 mins each with a cooldown break of several minutes in between. After completing the etching step, the sample was left in N-methyl-2-pyrrolidone (NMP) overnight to dissolve the rest of the ZEP photomask.

**Experimental setup**

A femtosecond laser system tunable across the mid-infrared range (based on a *Pharos* Yb:KGW regenerative amplifier, pumping an optical parametric amplifier *Orpheus* equipped with a difference-frequency-generation module *Lyra* – from *Light Conversion*) with pulse duration of approx. 150 fs was used to excite the samples. The repetition rate of the pump laser pulses impinging on the samples was 250 kHz. A telescope consisting of two $CaF_2$ lenses with focal lengths of 250 mm and 50 mm was utilized to reduce the beam waist at the sample plane to 230 μm (radius at $1/e^2$ of the maximum intensity). To achieve higher pump peak intensities, another set of $CaF_2$ lenses with focal lengths of 200 mm and 20 mm was used, further reducing the beam waist down to 105 μm. The spectra of the THG light emanating from the GSST samples were recorded using a monochromator (*Oriel* 74050– *Newport*). The spatial profile of the THG was detected using a thermo-electrically cooled InGaAs camera (*Xeva-1.7-320-Xenics*) with a spectral response covering the region 0.9 - 1.7 μm. To measure the TH power, we used a Ge-based high-gain detector (DET50B2: operating in the range 0.8 -1.8 μm) coupled to a lock-in amplifier. The detector was initially calibrated by setting the ultrafast laser wavelength at 1.5 μm.

**Simulations**

The simulation results presented in Fig. 2 (b)-(f) were obtained using a commercial software (CST Microwave Studio®) based on a finite integration technique (FIT). We computed the transmittance and reflectance of the GSST cylinder arrays on the $CaF_2$ substrate by sending a plane wave excitation onto a single cylinder confined in a unit cell with periodic boundary conditions in the directions (*x*, *y*) perpendicular to the propagation direction of the plane wave (*z*, normal incidence condition). The time-domain solver was used to perform the broadband simulations. The Poynting vector and EM field distribution maps were obtained by defining electric and magnetic field monitors at the wavelengths of interest.

## Acknowledgements

This work at INRS was supported by the Natural Sciences and Engineering Research Council of Canada (NSERC) through the Strategic and Discovery grant programs as well as by the Canada Research Chair Program. L. R. and R. M. would also like to acknowledge support from the Ministère de l'Économie et de l'Innovation du Québec (PSO-International). R. M. is affiliated to 5 as an adjoint faculty. M. Y. S., T. G., K. A. R., and J. H. would like to acknowledge funding from Defense Advanced Research Projects Agency Defense Sciences Office (DSO) Program: EXTREME Optics and Imaging (EXTREME) under Agreement No. HR00111720029. The authors also acknowledge characterization facility support provided by the Materials Research Laboratory at Massachusetts Institute of Technology (MIT), as well as fabrication facility support by the Microsystems Technology Laboratories at MIT and Harvard University Center for Nanoscale Systems.

## Author contributions

L. R., and J. H. conceived the study and supervised the project. K. R. prepared the GSST target materials. M. Y. S. fabricated the GSST films, as well as the metasurface, and characterized their linear properties. M. Y. S. and F. Y. conducted the numerical simulations. F. Y., R. P., and L. R. contributed to the characterization of the nonlinear properties of the fabricated samples and its interpretation. All the authors discussed the experimental results and contributed to the manuscript.